\documentclass{article} %
\usepackage{iclr2027_conference,times}

\usepackage{amsmath,amsfonts,bm}

\def\eqref#1{equation~\ref{#1}}

\def\1{\bm{1}}

\DeclareMathAlphabet{\mathsfit}{\encodingdefault}{\sfdefault}{m}{sl}
\SetMathAlphabet{\mathsfit}{bold}{\encodingdefault}{\sfdefault}{bx}{n}

\usepackage{amssymb}
\usepackage{microtype}
\usepackage{hyperref}
\usepackage{url}
\usepackage{booktabs}
\usepackage{graphicx}
\usepackage{amsmath}
\usepackage{algorithm}
\usepackage{algpseudocode}
\usepackage{subcaption}
\usepackage{wrapfig}
\usepackage{bbm}

\usepackage{amsthm}
\theoremstyle{definition}
\newtheorem{theorem}{Theorem}
\newtheorem{proposition}{Proposition}
\newtheorem{lemma}{Lemma}

\usepackage{enumitem}
\newenvironment{denseitemize}{
\begin{itemize}[topsep=0pt, partopsep=0pt, leftmargin=1.5em]
  \setlength{\itemsep}{2.5pt}
  \setlength{\parskip}{0pt}
  \setlength{\parsep}{0pt}
}{\end{itemize}}

\usepackage[most]{tcolorbox}

\tcolorboxenvironment{proposition}{
    colback=black!2,
    colframe=black!35,
    boxrule=0.5pt,
    arc=1mm,
    left=1.5mm,
    right=1.5mm,
    top=1mm,
    bottom=1mm,
    before skip=6pt,
    after skip=6pt
}
\usepackage{xspace}

\tcolorboxenvironment{theorem}{
    colback=black!2,
    colframe=black!35,
    boxrule=0.5pt,
    arc=1mm,
    left=1.5mm,
    right=1.5mm,
    top=1mm,
    bottom=1mm,
    before skip=6pt,
    after skip=6pt
}
\tcolorboxenvironment{lemma}{
    colback=black!2,
    colframe=black!35,
    boxrule=0.5pt,
    arc=1mm,
    left=1.5mm,
    right=1.5mm,
    top=1mm,
    bottom=1mm,
    before skip=6pt,
    after skip=6pt
}
\def\name{CacheFlow\xspace}

\definecolor{darkblue}{rgb}{0, 0, 0.5}
\hypersetup{colorlinks=true, citecolor=darkblue, linkcolor=darkblue, urlcolor=darkblue}

\title{CacheFlow: Efficient LLM Serving via \\ Automated 3D-Parallel KV Cache Restoration}

\author{
Sean Nian\textsuperscript{1*}\quad
Hanzhang Shen\textsuperscript{1*}\quad
Zhiyu Wu\textsuperscript{1}\quad
Jiahao Fang\textsuperscript{1}\quad
Qilong Feng\textsuperscript{2}\quad
Fan Lai\textsuperscript{1} \\
\textsuperscript{1}University of Illinois Urbana-Champaign \quad
\textsuperscript{2}Carnegie Mellon University \\
\texttt{\{snian2, shen74, zhiyuwu2, jiahaof3, fanlai\}@illinois.edu} \\
\texttt{qilongf@andrew.cmu.edu \quad \quad \quad \quad *Equal contribution} \\
}

\iclrfinalcopy
\hypersetup{
  pdftitle={CacheFlow: Efficient LLM Serving via Automated 3D-Parallel KV Cache Restoration},
  pdfauthor={Sean Nian; Hanzhang Shen; Zhiyu Wu; Jiahao Fang; Qilong Feng; Fan Lai},
  pdfsubject={LLM serving and KV cache restoration}
}
\begin{document}

\maketitle
\lhead{Preprint}

\begin{abstract}
KV cache restoration is becoming a major bottleneck in long-context LLM serving, including multi-turn conversations, agentic workloads, and RAG. Existing advances largely treat restoration as a coarse per-request choice between recomputing cached states and loading them from external storage (e.g., CPU memory or remote machines), overlooking structured parallelism within the model and resource contention across batched requests. Increasingly, modern hybrid architectures complicate restoration as their interleaved attention and recurrent layers introduce heterogeneous compute costs and state footprints. 
We present CacheFlow, which rethinks restoration as a multi-dimensional parallel execution problem. To our knowledge, CacheFlow is the first framework to jointly exploit
restoration parallelism across tokens, layers, and GPUs.
CacheFlow unifies token- and layer-wise restoration as a staircase partition
of the token--layer space, while lightweight boundary activations enable
concurrent restoration across model shards.
A batch-aware dynamic-programming planner jointly determines what to recompute
and what to load across requests and layer blocks, balancing GPU computation
against shared I/O resources. Across dense and hybrid models spanning single-digit to hundred-billion
parameters, diverse workloads, and hardware conditions, CacheFlow reduces restoration latency by 2.24$\times$--3.00$\times$ on average over existing advances, directly improving end-to-end serving latency by 1.64$\times$.

\end{abstract}

\section{Introduction}
Rapid adoption of LLM in interactive and agentic applications is driving a shift toward long-lived contexts. Modern workloads, including multi-turn conversations~\citep{contextpilot-mlsys26} and tool-augmented agents~\citep{agent-kvcache}, repeatedly reuse previous context. Efficiently restoring the KV cache---the cached intermediate state representing the previous context---has become a first-order serving challenge, as restoration lies directly on the critical path of Time-To-First-Token (TTFT), a key measure of responsiveness in user-facing systems~\citep{mooncake-fast25, jitserve-nsdi26}.

Existing KV cache restoration, however, remains fundamentally inefficient, incurring many seconds of latency for long-context requests. Existing advances often expose a coarse-grained dichotomy: they either recompute the cached prefix through prefill~\citep{pagedattention-sosp23,andes-arxiv24,continumm-arxiv25}, consuming scarce GPU compute, or load previously materialized KV states from CPU memory, SSDs, or remote machines~\citep{lmcache, echo-osdi26}, consuming I/O bandwidth. Neither resource is uniformly preferable: recomputation becomes increasingly expensive as the context grows, whereas loading is limited by storage and network bandwidth.

While existing hybrid approaches~\citep{cake-icml25,mooncake-fast25,hcache-eurosys25} partially overlap recomputation and cache transfer, they remain fundamentally request-centric and fail to capture two key properties and challenges of modern LLM serving. First, restoration costs are highly non-uniform across both tokens and layers. For full attention, recomputation cost grows superlinearly with sequence length, so the KV of later tokens is disproportionately more expensive to regenerate.
Modern hybrid architectures---such as Gated Delta Networks (GDN) in Qwen3.5~\citep{qwen3.5} and Kimi Linear~\citep{kimi-k3}---introduce an additional source of heterogeneity: full-attention and recurrent layers differ substantially in state size, recomputation cost, and the benefit of partial recomputation. As a result, a single recompute--load split for an entire request cannot capture the best tradeoff across the token--layer space. Second, production serving is inherently batched and distributed. Concurrent requests compete for shared GPU compute and I/O bandwidth, while model shards introduce cross-device execution dependencies. These effects create contention and stragglers, making independent per-request restoration decisions fundamentally insufficient for overall efficiency.

A decade of distributed deep learning has established a simple lesson: efficient model execution comes from exposing and exploiting its structural parallelism, from data parallelism to pipeline parallelism across devices~\citep{alpaserve-osdi23}. KV cache restoration, however, is still largely treated as a coarse recompute-or-load operation. Our key insight is that restoration exposes an analogous structure and can itself be parallelized across the token, layer, and device dimensions. Along the \emph{token} dimension, recomputing an early prefix while loading later KV states avoids the increasingly expensive tail of attention computation. Along the \emph{layer} dimension, recomputing lower layers while loading higher-layer states better amortizes execution overheads (e.g., weight movement). Yet, the preferred partition varies with requests, model architecture, and available resources. This motivates a unified view of cache restoration as a \emph{multi-dimensional parallel execution problem}.

We present \name, a KV cache restoration framework that realizes this view through coordinated 3D parallelism across tokens, layers, and GPUs. Rather than committing to a single token-wise or layer-wise split, \name generalizes both into a \emph{staircase partition} of the token--layer space, allowing different layer blocks to recompute different portions of the cached prefix while loading the remainder. At its core, a batch-aware dynamic-programming (DP) planner efficiently searches this space by retaining, for each achievable amount of avoided cache transfer, the minimum recomputation cost required to realize it. The same formulation extends across batched requests, enabling \name to jointly balance recomputation and I/O under shared serving resources.

\name further extends restoration along the device dimension, where recomputation would normally remain sequential across model shards because each downstream shard depends on activations from the preceding one. \name exploits lightweight \emph{boundary activations} at model's shard boundaries, allowing each GPU to independently execute the restoration plan for its local layers. Together, these mechanisms transform KV cache restoration from a coarse recompute-or-load decision into coordinated 3D parallel execution. Overall, 
we make the following key contributions:
\begin{denseitemize}
\item \textbf{The first multi-dimensional formulation of KV cache restoration.} To our knowledge, we are the first to formulate restoration as dependency-constrained partitioning over the token--layer space, capturing non-uniform recomputation costs, model architectures, and resource contention.

\item \textbf{A unified 3D-parallel KV cache restoration framework.} We introduce token-, layer-, and GPU-level parallelism through a batch-aware dynamic-programming planner that selects recomputation boundaries across requests and layers, and enables concurrent restoration across GPUs.

\item \textbf{An end-to-end system and comprehensive evaluation.} We implement CacheFlow atop vLLM and LMCache, transparently supporting
existing LLM applications.
We evaluate it from single-digit-billion dense models to
hundred-billion-parameter sparse hybrid models (e.g., Qwen3.5-122B-A10B), across realistic chatbot and
agentic workloads and diverse hardware and I/O conditions.
CacheFlow reduces TTFT by 2.24$\times$--3.00$\times$ on average and TTLT by 1.64$\times$.
\end{denseitemize}

\section{Related Work and Motivation}
\label{sec:background}
LLM serving consists of two sequential stages: a \emph{prefill} stage that processes the input context, and a \emph{decode} stage that autoregressively generates output tokens~\citep{agrawal2024taming, jitserve-nsdi26}.
Prefill produces the key and value (KV) representations reused by subsequent attention computations.

\textbf{KV cache optimization.}
KV cache avoids repeatedly processing context, benefiting multi-turn conversations~\citep{deshpande-etal-2025-multichallenge}, RAG~\citep{cacheblend-eurosys25}, and agentic workflows~\citep{merrill2026terminal}.
However, cache size grows linearly with context length and dominates GPU memory usage.
Previous work reduces KV cache memory through quantization~\citep{zandieh2026turboquant, shen2026triaxialkv, liu2026pmkvq, hooper2024kvquant, liu2024kivi}, token eviction~\citep{rkv-neurips25, cai2024pyramidkv, bumblebee-colm24, zhang2023h2o}, or a combination of both~\citep{diffkv-sosp25}.
Other systems offload cached states to lower memory tiers~\citep{xie2026strata, echo-osdi26, lmcache, mooncake-fast25}, or exploit the workflow structure to guide cache retention and prefetching~\citep{continumm-arxiv25, kvflow-neurips25}.
When a needed prefix's KV cache is no longer GPU-resident, it must be restored by loading a saved copy from lower memory tiers or recomputing it~\citep{jitserve-nsdi26, pagedattention-sosp23}, placing restoration on the critical path to the first output token.

\textbf{KV cache restoration.}
KV recomputation consumes GPU compute, while cache loading uses I/O bandwidth, motivating hybrid approaches that overlap the two operations.
Cake~\citep{cake-icml25} recomputes early tokens while loading the cached KV states of later tokens.
Similarly, KVPR~\citep{jiang-etal-2025-kvpr} overlaps partial KV recomputation with transfer of the remaining KV during offloaded decoding.
Hybrid approaches also appear in CacheBlend~\citep{cacheblend-eurosys25} and eLLM~\citep{ellm2026chen}, targeting RAG contexts and partial-token caching respectively.
These systems have shown the benefits of combining recomputation and transfer.
CacheFlow extends this approach through joint token–layer planning across requests and concurrent restoration across devices.

\textbf{Limitations of prior advances.}
The relative efficiency of recomputation and cache loading varies with context length, model architecture, hardware, and available I/O bandwidth. Recent hybrid architectures, as in Gated Delta Networks (GDNs)~\citep{yang2025gated} in Qwen3.5~\citep{qwen3.5} and Kimi
    Linear~\citep{team2025kimi} in Kimi-K3~\citep{kimi-k3}, make this tradeoff substantially harder.
    Existing restoration techniques fail to capture two key properties in modern
    settings:

\begin{denseitemize}
    \item \emph{Restoration costs vary across tokens and layer types.}
    Full-attention layers maintain token-proportional KV states, whereas
    recurrent layers maintain compact fixed-size states, leading to very
    different recomputation--loading tradeoffs across model depth.
    Moreover, partial recomputation directly avoids KV transfer in
    full-attention layers, but does not eliminate the final recurrent state
    required for decoding.
    Existing schemes such as Cake and HCache~\citep{hcache-eurosys25} do not directly
    generalize to recurrent states, and/or lose their storage advantage as hidden activations
    are comparable to recurrent states.
    A single restoration policy across the model becomes insufficient.

    \item \emph{Per-request optimization ignores system-wide resource sharing.}
    Production serving batches requests that contend for shared
    GPU compute and I/O bandwidth, while distributed execution introduces
    dependencies across model shards.
    Optimizing each request independently creates resource
    imbalance and stragglers even when each local restoration decision appears
    efficient.
\end{denseitemize}

CacheFlow rethinks restoration as a broader joint-optimization problem across tokens, layers, devices, and requests, accounting for heterogeneous costs across layers, and planning for a batch.

\section{CacheFlow: Automated 3D-Parallel KV Cache Restoration}

\label{sec:design}

\begin{figure}[t]
    \begin{center}
        \includegraphics[width=.9\linewidth]{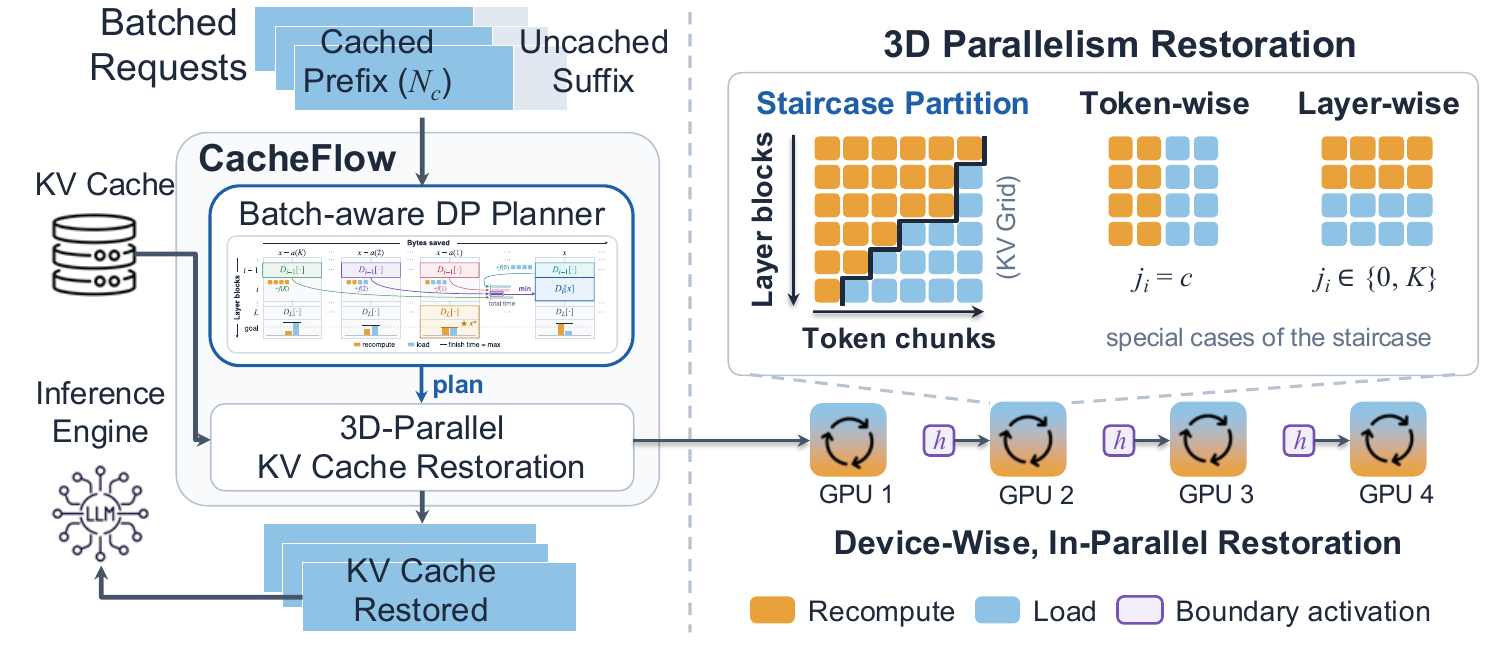}
    \end{center}
    \vspace{-.4cm}
    \caption{\textbf{\name overview.} The batch-aware DP planner partitions each request's token--layer KV grid into a staircase to maximize computation and I/O overlap across layers, tokens, and devices.}
    \label{fig:system_overview}
    \vspace{-.4cm}
\end{figure}

We present \name, a KV cache restoration framework that rethinks restoration as a multi-dimensional parallel execution problem.
Our design is grounded in a unifying principle: \emph{KV cache restoration can exploit parallelism along the token, layer, and device dimensions.}
As shown in Figure~\ref{fig:system_overview}, given a request with a cached prefix of $N_c$ tokens, \name aims to minimize its KV restoration latency by jointly coordinating recomputation and cache loading.

We begin with introducing 3D restoration parallelism across tokens, layers, and devices (\S\nobreak\ref{sec:intra-request}).
Token-wise and layer-wise restoration correspond to vertical and horizontal partitions of the token--layer grid, exposing different compute--I/O tradeoffs that depend on request length and model architecture, particularly for modern hybrid-attention models.
We extend these strategies across devices so that GPUs can restore their local states concurrently.
Then we generalize these partitions into a planning problem over the token--layer space and solve it for individual requests and then for a batch (\S\nobreak\ref{sec:planned-partitioning}).

\subsection{3D Restoration Parallelism}
\label{sec:intra-request}

Existing systems (e.g., vLLM and LMCache) typically restore a cached prefix in one of two ways:
\emph{recompute} its KV through forward execution, consuming GPU compute,
or \emph{load} previously materialized KV from storage, consuming I/O bandwidth.
Relying exclusively on either resource leaves the other underutilized.
Rather than making this choice for an entire request, \name exposes
opportunities for recomputation and loading to proceed concurrently.
We first exploit the token and layer dimensions within a model shard, and
then extend restoration across devices.

\begin{wrapfigure}{r}{2in}
  \centering
  \vspace{-\intextsep}
  \includegraphics[width=.95\linewidth]{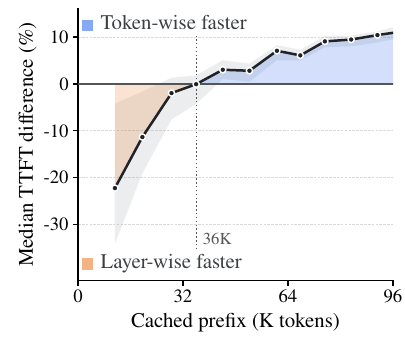}
  \vspace{-.3cm}
  \caption{Paired per-request TTFT gap between token-wise and
  layer-wise restoration on Llama-3.1-8B. Positive means token-wise is
  faster.}
  \label{fig:token-layer-crossover}
  \vspace{-.4cm}
\end{wrapfigure}

\textbf{Token-wise parallelism.}
Consider a request's token--layer KV grid (Figure~\ref{fig:system_overview}). By dividing the prefix of $N_c$ tokens into $K=\lceil N_c/C\rceil$ chunks indexed $0, 1, \ldots, \lceil N_c / C \rceil - 1$, we can recompute chunks from the first chunk of the sequence while loading the KV cache from the end.
The two processes execute concurrently until they meet at an intermediate chunk.
Chunking serves three purposes.
First, it exposes fine-grained scheduling units, allowing recomputation to
stop as soon as it meets the concurrently loaded suffix, thus avoiding redundancy.
Second, choosing $C$ to align with today's efficient GPU execution and scheduling granularities
(e.g., 512 tokens per chunk in FlashAttention) preserves high recomputation
efficiency~\citep{flashattention, cake-icml25}.

Third, the opposite traversal directions are intentional.
Under full attention, the marginal cost of recomputing a token increases with
its position because it attends to an increasingly long prefix; theoretically,
the cumulative recomputation cost grows quadratically with sequence length~\citep{Vaswani+2017}.
In contrast, the KV state transferred per token is approximately constant.
Therefore, token-wise restoration avoids recomputing the
increasingly expensive tail.

\textbf{Layer-wise parallelism.}
The same recomputation--loading overlap can be exploited along the layer dimension.
Here, recomputation proceeds forward from the first layer, while cached states are loaded backward from the last layer.
For each layer, the states for the entire cached prefix are recomputed or loaded before advancing to the next layer.
The two operations run concurrently and meet at a layer boundary.
Thus, layers before the boundary are recomputed for the full cached prefix.

Layer-wise restoration exposes a complementary tradeoff.
Because each selected layer recomputes the entire cached prefix, it incurs the
expensive attention tail that token-wise restoration avoids.
In return, it better amortizes fixed execution overheads: token-wise
restoration repeatedly incurs kernel-launch, synchronization,
and weight-movement overheads across chunks.
Layer-wise restoration pays these costs once per layer.
Figure~\ref{fig:token-layer-crossover} illustrates the trade-off:
layer-wise restoration can be preferable for shorter prefixes,
whereas token-wise restoration becomes increasingly attractive as attention
cost grows with context length.

\textbf{Device-wise parallelism.}
Device parallelism is complementary to these token--layer decisions.
Practical serving often partitions groups of model layers across
devices, either directly or as part of a hybrid parallel deployment~\citep{alpaserve-osdi23}.
While token-wise and layer-wise restoration expose parallelism, forward recomputation across model shards remains sequential: recomputation on a downstream shard cannot begin until the preceding shard produces its input activations.

\name removes this dependency using lightweight \emph{boundary activations}.
At each shard boundary, we cache the activation that forms the input to the
first layer of the downstream shard.
These activations are typically much smaller than the aggregate cached states
of all layers assigned to the shard, yet provide the input required to begin
its local forward execution.
During restoration, each GPU retrieves its boundary activation and executes
the token--layer restoration plan for its local layers independently of live
upstream execution.
Thus, model shards depend on cached boundary states rather than one another,
transforming restoration from a sequential pipeline into concurrent
shard-local execution. 
This mechanism composes naturally with both token-wise and layer-wise
restoration and is compatible with existing tensor and expert parallelisms too (\S\ref{eval:ablation}).
Model shards can concurrently recompute or load their local states according to
the plan derived in Section~\ref{sec:planned-partitioning}.

\subsection{Batch-Aware Planning over the Token--Layer Space}
\label{sec:planned-partitioning}

The preceding mechanisms expose restoration parallelism, but leave open how
much of each request should be recomputed versus loaded, especially under batch execution.
The best choice varies with context length, model architecture, batch
composition, and available compute and I/O bandwidth.
Moreover, modern hybrid architectures interleave full-attention layers with
recurrent or linear-attention layers whose state footprints and recomputation
costs differ substantially. At 128K context, full-attention layers have
125$\times$ and 250$\times$ larger state footprints than a recurrent checkpoint
in Qwen3.5-35B-A3B and Qwen3.5-9B, respectively, while measured per-layer
recomputation time differs by 2.6--4.5$\times$ on an H100. We next formulate restoration as partitioning the token--layer space.

\textbf{Restoration as a staircase partition.}
Consider a model consisting of $L$ layer blocks.
For a hybrid architecture, a block contains the recurrent/linear-attention
layers preceding a full-attention layer, such as 3:1 in Qwen3.5.
We use the block as the planning unit because recomputing its KV requires the activations produced by the preceding layers within that
block.
A dense transformer is a special case in which a block contains one
full-attention layer. Importantly, our formulation can be extended to heterogeneous layer blocks and architectures with multi-token prediction (MTP) modules by incorporating their respective computation costs, state sizes, and execution dependencies.

For a request with $K$ cached chunks, let $j_i \in \{0, \ldots, K\}$ denote the number of chunks recomputed in layer block $i$.
The full-attention KV states for the remaining $K-j_i$ chunks are loaded from external storage. 
A valid restoration plan
$\mathbb{J}=(j_1,\ldots,j_L)$ must respect forward dependencies:
if layer block $i+1$ recomputes a token chunk, block $i$ must also recompute that
chunk to produce its input activation. Therefore,
\begin{equation}
    K \ge j_1 \ge j_2 \ge \cdots \ge j_L \ge 0.
\label{eq:constraint}
\end{equation}
Geometrically, these boundaries form a \emph{staircase} over the token--layer
grid. A constant boundary $j_1=\cdots=j_L=c$ recovers token-wise restoration,
whereas setting the first several blocks to $K$ and the remaining blocks to
$0$ recovers layer-wise restoration. 

\paragraph{Modeling recomputation and transfer.}
Let $f(j)$ denote the time to recompute the first $j$ chunks through one
layer block, with $f(0)=0$.
Let $S_{\mathrm{attn}}$ denote the full-attention KV size per chunk,
$S_{\mathrm{rec}}$ the final recurrent-state size of the block, and $B$ the
available loading bandwidth.
Recomputing $j$ chunks in a block requires loading $d(j) = (K-j)S_{\mathrm{attn}} + \mathbbm{1}_{\{j < K\}}S_{\mathrm{rec}}$ bytes.
The first term counts the remaining KV chunks to load, and the second reflects that partial recomputation does not reconstruct the final
recurrent state, unless the entire prefix
is recomputed.
For a plan $\mathbb{J}=(j_1, \ldots, j_L)$, total computation takes $T_{\mathrm{comp}}(\mathbb{J})=\textstyle\sum_{i=1}^{L} f(j_i)$ while loading takes $T_{\mathrm{load}}(\mathbb{J})=\textstyle\sum_{i=1}^{L}d(j_i)/B$.
These two run concurrently, so restoration finishes when the slower
resource completes. Our objective is:
\begin{equation}
    \min_{K \ge j_1 \ge \cdots \ge j_L \ge 0} \left\{
    \max(
        T_{\mathrm{comp}}(\mathbb{J}),
        T_{\mathrm{load}}(\mathbb{J}))
    \right\}.
    \label{eq:restoration-objective}
\end{equation}
Enumerating all $(K+1)^L$ restoration plans is intractable. 
Our key insight is that for repeated blocks, choosing boundary $j$ incurs the same recomputation cost
$f(j)$ and avoids the same cache transfer $a(j)$ regardless of the block
index.
Hence, for any set of selected boundaries
$\{j_1,\ldots,j_L\}$, both
$\sum_i f(j_i)$ and $\sum_i a(j_i)$ depend only on the multiset of boundary
values, not on their assignment to model depth. Thus, permuting these boundaries leaves the restoration objective
unchanged.
We can thus sort them in non-increasing order,
$j_1 \ge \cdots \ge j_L$, to satisfy the forward dependencies in
Equation~(\ref{eq:constraint}) without sacrificing optimality.
These observations motivate a dynamic program (DP) that records the \emph{minimum recomputation time for each achievable amount of avoided cache transfer}.

\paragraph{Dynamic programming over KV grid.}
\begin{figure}[t]
    \centering
        \includegraphics[width=.95\linewidth]{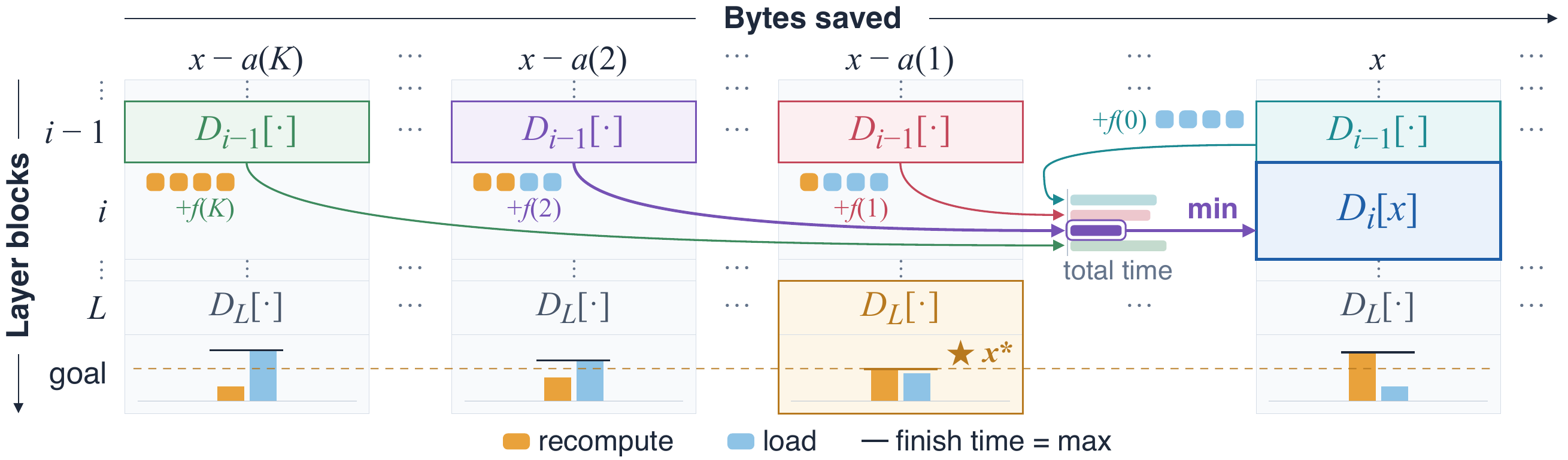}
    \vspace{-.2cm}
    \caption{\textbf{Illustration of our DP planner.} $D_i[x]$ takes the minimum of $D_{i-1}[x-a(j)] + f(j)$, where boundary $j$ recomputes $j$ chunks in block $i$ and loads the rest. For each final state $D_L[x]$, since recomputation and loading overlap, we pick the $x^\star$ minimizing $\max\{D_L[x], (A_0-x)/B\}$.}
    \label{fig:dp}
    \vspace{-.4cm}
\end{figure}

We next introduce the DP formulation. As shown in Figure~\ref{fig:dp}, for a block choosing recomputation boundary $j$, the avoided cache transfer is
$
    a(j)
    =
    jS_{\mathrm{attn}}
    +
    \mathbbm{1}_{\{j=K\}}S_{\mathrm{rec}}
$,
where the first term corresponds to the $j$ recomputed KV chunks, and the
second captures the recurrent state, whose loading is avoided only when the
entire prefix is recomputed. 
Let $D_i[x]$ denote the minimum recomputation time for the first $i$ layer
blocks while avoiding $x$ bytes of cache transfer in total.
Starting from $D_0[0]=0$ and $D_0[x]=+\infty$, we compute
\begin{equation}
    D_i[x]
    =
    \min_{j\in\{0,\ldots,K\}}
    \left\{
        D_{i-1}[x-a(j)] + f(j)
    \right\},
    \label{eq:single-request-dp}
\end{equation}
with unavailable states (e.g., due to eviction) and $x-a(j)<0$ assigned infinite cost.

After processing $L$ layer blocks, each reachable state $D_L[x]$ represents the minimum recomputation time required to avoid $x$ units of state transfer.
Without recomputation, the total amount of state to load is
$A_0=L(KS_{\mathrm{attn}}+S_{\mathrm{rec}})$.
As recomputation and loading execute concurrently, we select:
\begin{equation}
    x^\star
    =
    \operatorname*{arg\,min}_{x}
    \max\left\{
        D_L[x],
        \frac{A_0-x}{B}
    \right\}.
    \label{eq:dp-selection}
\end{equation}

\textbf{Joint planning across a batch.}
Optimizing requests independently can produce poor resource allocation when
they compete for the same GPU compute and I/O bandwidth.
Our formulation extends directly to a batch by treating every
\emph{request--block pair} as a planning item.

For a batch $\mathcal{R}$, let $M=|\mathcal{R}|L$ and let
$(r_m,i_m)$ denote the $m$-th request--block pair.
If request $r$ contains $K_r$ cached chunks, choosing boundary $j$ avoids
$
    a_r(j)
    =
    jS_{\mathrm{attn}}
    +
    \mathbbm{1}_{\{j=K_r\}}S_{\mathrm{rec}}
$
bytes of transfer.
Let $D_m[x]$ denote the minimum aggregate recomputation time after considering
the first $m$ request--block pairs while avoiding $x$ bytes of transfer.
The recurrence becomes
\begin{equation}
    D_m[x]
    =
    \min_{j\in\{0,\ldots,K_{r_m}\}}
    \left\{
        D_{m-1}[x-a_{r_m}(j)] + f_{r_m}(j)
    \right\}.
    \label{eq:batch-dp}
\end{equation}
The same final balancing step selects the batch-wide operating point between
recomputation and cache loading.
After backtracking, we group the recovered boundaries by request and reorder
each request's boundaries independently to obtain a dependency-valid
restoration plan.

\textbf{Putting planning into practice.}
We make the planner lightweight in three ways.
First, rather than tracking cache traffic at byte granularity, we quantize
transfer savings at the same chunk granularity used for restoration
(e.g., 512-token chunks). 
This reduces the DP state space substantially while retaining the granularity
at which \name can actually partition recomputation and loading.
Recurrent-state sizes are mapped to the nearest number of chunk-equivalent
transfer units.

Second, recomputation costs need not be modeled analytically.
For a fixed model and hardware, this cost is largely
determined by the token length, agnostic to content, and LLM
serving naturally executes requests across a wide range of prefix lengths.
Therefore, \name maintains an online latency profile $f(j)$ by collecting
recomputation times from recent serving executions and continuously updating
the corresponding profile entries.
Unobserved chunk counts are estimated by interpolation. 
These measurements are obtained from work the system already
executes. Our evaluations show that such collection maintains high accuracy, introduces little planning overhead, and
naturally adapts to changes in hardware and execution conditions (\S\ref{eval:ablation}).

Third, \name naturally incorporates the uncached suffix without expanding the DP state
space.
Suffix tokens have no cached state and are always recomputed.
For block $i$, suffix prefill can begin only after all prefix states required
by that block---its full-attention KV and final recurrent state---have been
restored, whether through recomputation or loading.
Once available, suffix computation is independent of how the prefix was
restored, so its cost is fixed for a given request and can be obtained from
the same latency profile.
With block-ordered loading, suffix prefill in earlier blocks can overlap
cache loading for later blocks.
Thus, if $T_{\mathrm{suf}}$ denotes the suffix-prefill time per block,
Equation~\ref{eq:dp-selection} becomes
$
x^\star
=
\operatorname*{arg\,min}_{x}
\max\left\{
    D_L[x] + LT_{\mathrm{suf}},
    (A_0-x)/B + T_{\mathrm{suf}}
\right\}$,
where the compute term covers all $L$ blocks' suffix prefill, and the loading term only the last block's, which must wait for all loads.
The DP recurrence remains unchanged. Appendix~\ref{app:dp} provides the analysis.

\textbf{Approximation guarantee and scaling capability.}
We next quantify \name's approximation gap relative to an idealized optimum
$T^*$, where every layer may independently choose an arbitrary token-level
restoration boundary (proof details in Appendix~\ref{app:dp}):

\begin{proposition}[Restoration optimality gap]\label{prop:gap}
Consider a request with $N_c$ cached tokens in $L$ layer blocks and chunk size $C$. Let $\tilde J$ be the plan returned by the DP, with makespan
$T(\tilde J)=\max\{T_{\mathrm{comp}}(\tilde J),T_{\mathrm{load}}(\tilde J)\}$
as in Equation~\ref{eq:restoration-objective}, and let $e$ be the rounding error of $a(K)$. Assume that the plan's loading
takes at least as long as the suffix prefill, the usual case since cached
prefixes are typically much longer than new suffixes, so that $T(\tilde J)$
is the plan's makespan. Let $\kappa$
lower-bound the per-token recomputation time of one block, and let
$\gamma=1+s/(B\kappa)$. Then
\begin{equation}\label{eq:gap}
T(\tilde J)-T^*
\le\frac{S_{\mathrm{rec}}+Ls(C-1)+L|e|}{B}
\le\gamma\,T^*\Bigl(\frac{S_{\mathrm{rec}}}{LN_cs}+\frac{C-1}{N_c}
+\frac{|e|}{N_cs}\Bigr).
\end{equation}
\end{proposition}
The three terms capture block-level planning, chunking, and transfer
quantization, whose relative impact vanishes as $O(1/N_c)$.
For Qwen3.5-9B at 10\,Gbps with $C=512$, the theoretical gap is at most 19.5\,ms,
corresponding to 6.52\%, 3.26\%, and 1.63\% of $T^*$ for 32K-, 64K-, and
128K-token prefixes, respectively.
For full attention at token granularity
($S_{\mathrm{rec}}=0$, $C=1$, $e=0$), the gap vanishes. Further, device-wise parallelism provides a complementary near-linear scaling benefit
with the number of model shards (Appendix~\ref{app:speedup}). We empirically validate this
scaling capability in \S\ref{eval:ablation}.

\section{Evaluation}
\label{sec:eval}

\subsection{Experimental Setup}
\label{sec:setup}

\noindent \textbf{Models and Workloads.}
We implemented CacheFlow atop vLLM and LMCache. We evaluated \name on Llama-3.1-8B~\citep{grattafiori2024llama} (full attention), Qwen3.5-9B~\citep{qwen3.5} (hybrid attention),  Qwen3.5-35B-A3B (MoE, hybrid attention), and larger Qwen3.5-122B-A10B.
They expose different state structures and compute--I/O tradeoffs.
Our workloads span three realistic datasets:
(i) \emph{LMSYS-Chat}~\citep{lmsys-chat}, which consists of real ChatGPT multi-turn conversational traces where successive turns share long prefixes;
(ii) \emph{SWE-Bench}~\citep{swe-bench}, an agentic coding benchmark with repeated tool invocations over shared repository contexts.
(iii) \emph{OSWorld 2.0}~\citep{yuan2026osworld2}, a computer-use agentic benchmark.

\noindent \textbf{Hardware and Network Conditions.} 
We conducted experiments on NVIDIA A100s and H100s, covering both single-GPU and multi-GPU deployments.
We evaluate different practical bandwidth conditions of 80\,Gbps, 40\,Gbps, and 10\,Gbps, corresponding to typical Infiniband/RoCE speed~\citep{cake-icml25}, Lambda Lab SSD read speed, and Amazon Cloud inter-node bandwidth~\citep{cachegen-sigcomm24}, respectively. Additionally, we evaluate \name in online serving using Mooncake arrival timing~\citep{mooncake-fast25}.
By default, we use 10\,Gbps, with ablation studies in Section~\ref{eval:ablation}.

\noindent \textbf{Baselines.} 
We compare \name with three advances:

\begin{denseitemize}
    \item \emph{vLLM} v0.29.0~\citep{vllm-github}: it uses recomputation-only as its default policy.
    
    \item \emph{LMCache} v0.5.5rc5~\citep{lmcache-github}: KV cache loading without recomputation, representing the I/O-bound extreme and the state-of-the-art offloading system.
    
    \item \emph{Cake}~\citep{cake-icml25}: a strong prior token-wise hybrid restoration approach that partitions each request's cached prefix, recomputing its leading portion while loading its trailing portion.
\end{denseitemize}

\noindent \textbf{Metrics.} We aim to minimize \emph{Time-To-First-Token} (TTFT), which directly captures the latency impact of KV cache restoration on user-perceived responsiveness, and \emph{Time-To-Last-Token} (TTLT) in online deployments. We also report GPU compute utilization and I/O utilization during restoration. 

\subsection{End-to-End Performance}
\label{eval:e2e}

\begin{figure}[t]
    \centering
        \includegraphics[width=.9\linewidth]{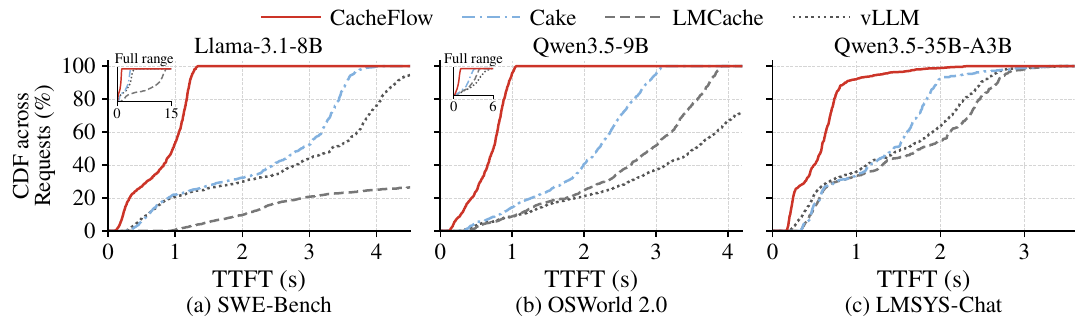}
    \vspace{-.3cm}
    \caption{CacheFlow achieves lower serving latency than existing advances.}
    \label{fig:e2e-ttft-cdf}
    \vspace{-.4cm}
\end{figure}

\textbf{\name reduces serving latency.} 
Figure~\ref{fig:e2e-ttft-cdf} shows the TTFT distribution across requests for LMSYS-Chat, SWE-Bench, and OSWorld 2.0. Across all workloads, CacheFlow consistently left-shifts the CDF, indicating lower latency at every percentile. On average, CacheFlow achieves a 2.24$\times$--3.00$\times$ reduction in TTFT compared to existing advances. The gains are most pronounced on LMSYS-Chat and SWE-Bench, which feature longer contexts and thus incur higher KV restoration costs.  
Notably, the improvement extends to the tail: CacheFlow achieves 1.77$\times$--3.02$\times$ and 1.48$\times$--2.95$\times$ speedups at p95 and p99, respectively. In \S\ref{eval:ablation}, we show that \name improves time-to-last-token (TTLT) performance by 1.64$\times$. 

\begin{figure}[t]
\centering
\captionsetup{hypcap=false}
\begin{minipage}[t]{0.23\linewidth}
\vspace{0pt}
        \centering
        \includegraphics[width=\linewidth]{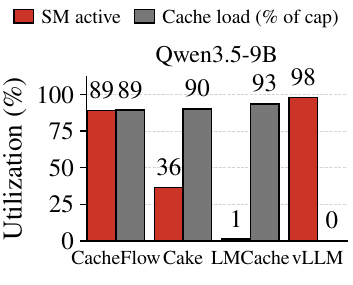}
        \vspace{-5mm}
        \captionof{figure}{Breakdown of Resource utilization.}
        \label{fig:utilization-breakdown}
    \end{minipage}
\hfill
\begin{minipage}[t]{0.23\linewidth}
\vspace{0pt}
\centering
\captionsetup{hypcap=false}
\includegraphics[width=\linewidth]{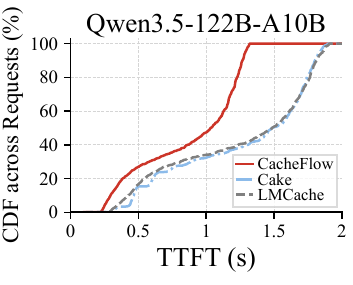}
\vspace{-5mm}
\captionof{figure}{Large-scale MoE serving.}
\label{fig:large-moe-tp8-pp1}
\end{minipage}
\hfill
\begin{minipage}[t]{0.23\linewidth}
\vspace{0pt}
        \centering
        \includegraphics[width=\linewidth]{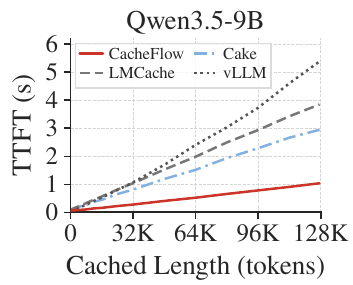}
        \vspace{-5mm}
        \captionof{figure}{TTFT by input length.}
        \label{fig:ablation-input-length}
\end{minipage}
\hfill
\begin{minipage}[t]{0.28\linewidth}
\vspace{0pt}
        \centering
        \includegraphics[width=\linewidth]{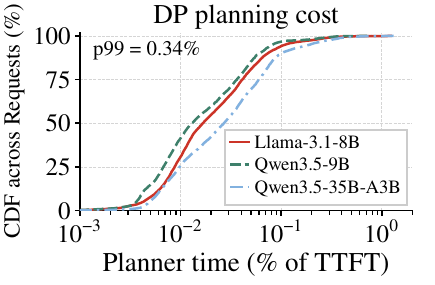}
        \vspace{-5mm}
        \captionof{figure}{Planning is 0.34\% of TTFT at p99.}
        \label{fig:planner-cost}
    \end{minipage}
\vspace{-.5cm}
\end{figure}

\textbf{\name improves resource utilization.} 
Figure~\ref{fig:utilization-breakdown} reports average GPU and I/O utilization during KV cache restoration. LMCache saturates I/O bandwidth but achieves only 1\% GPU utilization, as KV transfers throttle execution. Conversely, vLLM is compute-bound with 98\% GPU utilization but idle I/O. In contrast, \name achieves 89\% GPU and 89\% I/O utilization by effectively overlapping recomputation and I/O.

\textbf{\name is compatible with model parallelisms.}
We further evaluate \name on Qwen3.5-122B-A10B using eight H100 GPUs with TP=8/PP=1 and TP=4/PP=2, with expert parallelism (EP) disabled and enabled. Figure~\ref{fig:large-moe-tp8-pp1} and Appendix~\ref{app:large-moe} show TTFT distributions on SWE-Bench with batch size 8 and 10\,Gbps I/O bandwidth per worker. All methods use the same parallel configuration within each panel. Across these settings, \name achieves 1.21$\times$--1.43$\times$ speedups in mean TTFT.

\subsection{Performance Breakdown and Ablation Study}
\label{eval:ablation}

\textbf{Breakdown by request lengths.}
Figure~\ref{fig:ablation-input-length} breaks down \name's performance across different request length regimes on Qwen3.5-9B with OSWorld 2.0. We notice that recomputation-based methods (e.g., vLLM) incur substantially higher latency for long requests. In contrast, \name scales more gracefully by overlapping recomputation and I/O. Overall, \name achieves 3.00$\times$, 3.96$\times$, and 4.82$\times$ speedups in mean TTFT compared to Cake, LMCache, and vLLM, respectively.

\begin{figure*}[t]
\captionsetup{hypcap=false}
\centering
\begin{minipage}[t]{0.49\textwidth}\vspace{0pt}
    \centering
    \includegraphics[width=\linewidth]{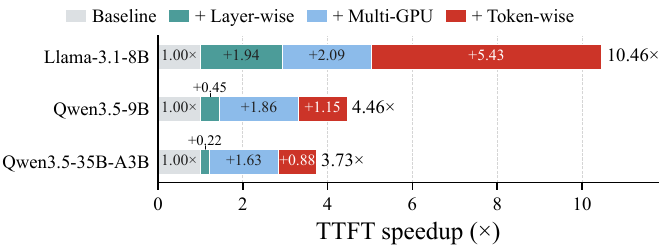}
    \vspace{-.6cm}
    \caption{Performance breakdown.}
    \label{fig:restoration-breakdown}
\end{minipage}
\hfill
\begin{minipage}[t]{0.24\linewidth}
\vspace{0pt}
        \centering
        \includegraphics[width=\linewidth]{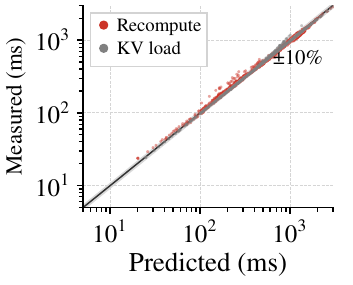}
        \captionof{figure}{Predicted vs. measured latency.}
        \vspace{-5mm}
        \label{fig:planner-accuracy}
    \end{minipage}
\hfill
\begin{minipage}[t]{0.24\linewidth}
\vspace{0pt}
        \centering
        \includegraphics[width=\linewidth]{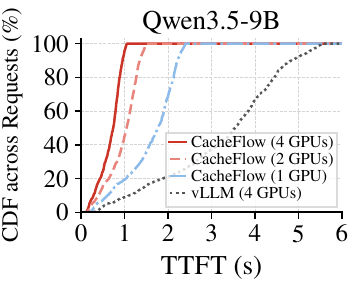}
        \vspace{-5mm}
        \captionof{figure}{Strong scaling in \name.}
        \label{fig:ablation-multi-gpu}
    \end{minipage}
\vspace{-.3cm}
\end{figure*}

\textbf{Performance breakdown by components.}
Figure~\ref{fig:restoration-breakdown} breaks down \name's end-to-end TTFT on H100 GPUs with PP=4 and 10\,Gbps I/O bandwidth per worker by progressively enabling layer-wise restoration, multi-GPU execution, and token-wise restoration. We observe that these mechanisms provide complementary benefits: together, they improve mean TTFT by 10.46$\times$, 4.46$\times$, and 3.73$\times$ over the single-GPU load-only baseline on Llama-3.1-8B, Qwen3.5-9B, and Qwen3.5-35B-A3B, respectively. These results highlight the effectiveness of combining restoration parallelism.

\textbf{DP planner cost and accuracy.}
During a sweep of three models and three bandwidths, the DP planner costs 0.019\% of TTFT at the median, and 0.34\% at p99 (Figure~\ref{fig:planner-cost}).
Our DP is not only fast but also accurate.
Figure~\ref{fig:planner-accuracy} shows 92\% of recompute predictions and 98\% of load predictions land within 10\% of measured latency.
We also measured how evenly the plan splits the work.
Recompute and load finish within 10\% of each other on 95\% of restorations.

\textbf{Impact of GPU counts.} 
We break down our 3D parallelism design by varying the GPU count from one to two and four GPUs. Figure~\ref{fig:ablation-multi-gpu} shows that average TTFT decreases from 1.59\,s to 0.96\,s and 0.67\,s, respectively (a 2.36$\times$ speedup with four GPUs). Notably, even with a single GPU, \name still outperforms vLLM by 51\%, highlighting the effectiveness of our batch-aware DP design. 

\begin{figure*}[t]
\captionsetup{hypcap=false}
\centering
\begin{minipage}[t]{0.24\linewidth}
\centering
\vspace{0pt}
\includegraphics[width=\linewidth]{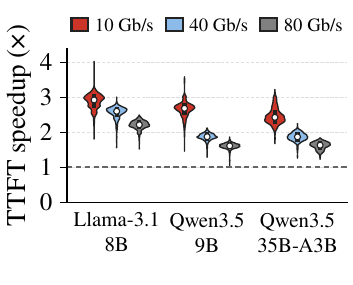}
\vspace{-5mm}
\captionof{figure}{Impact of I/O bandwidth.}
\label{fig:ablation-io-bandwidth}

\end{minipage}
\hfill
\begin{minipage}[t]{0.24\linewidth}
\vspace{0pt}
\centering

\includegraphics[width=\linewidth]{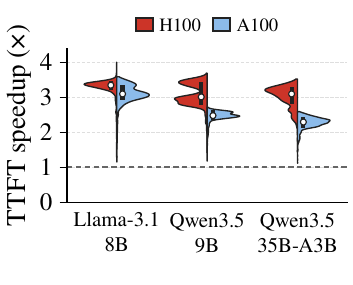}
\vspace{-5mm}
\captionof{figure}{Impact of GPU hardware.}
\label{fig:ablation-gpu-hardware}

\end{minipage}
\hfill
\begin{minipage}[t]{0.49\textwidth}\vspace{0pt}
    \centering
    \begin{minipage}[b]{0.49\linewidth}
        \centering
        \includegraphics[width=\linewidth]{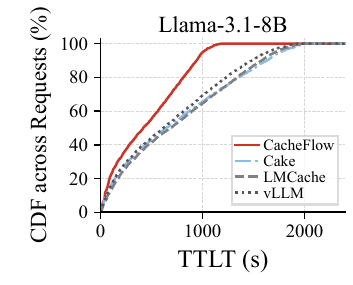}
    \end{minipage}
    \hfill
    \begin{minipage}[b]{0.49\linewidth}
        \centering
        \includegraphics[width=\linewidth]{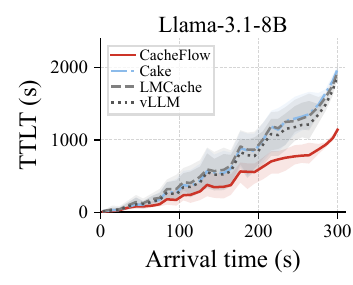}
    \end{minipage}
    \vspace{-.6cm}
    \caption{Online serving deployments.}
    \label{fig:ablation-online-comparison}
    \label{fig:online-burst-backlog}
\end{minipage}
\vspace{-.5cm}
\end{figure*}

\textbf{Impact of I/O bandwidth.} We evaluate \name under different I/O bandwidths from our end-to-end evaluations: 40\,Gbps and 80\,Gbps on our H100 cluster. Figure~\ref{fig:ablation-io-bandwidth} shows that \name consistently improves mean TTFT, yielding 2.63$\times$ and 2.40$\times$ speedups over Cake at 40\,Gbps and 80\,Gbps, respectively. These gains stem from \name's profile-guided DP design, which balances recomputation and I/O across tokens and layers. Under higher bandwidth, more KV cache can be loaded in parallel, further reducing restoration latency.

\textbf{Impact of GPU hardware.} 
Figure~\ref{fig:ablation-gpu-hardware} further ablates our previous setting by varying the hardware from H100s to A100s on Qwen3.5-35B-A3B with OSWorld 2.0, under 10\,Gbps KV cache I/O transfer per worker. We observe that \name consistently outperforms Cake, achieving 2.69$\times$ and 2.07$\times$ speedups in mean TTFT on H100 and A100, respectively. 

\begin{wrapfigure}{R}{0.32\textwidth}
\vspace{-1cm}
\centering
\includegraphics[width=\linewidth]{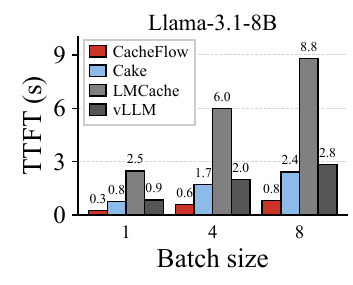}
\vspace{-.7cm}
\caption{Impact of batch size.}
\label{fig:ablation-batch-size}
\vspace{-.6cm}
\end{wrapfigure}
\textbf{Online serving deployments.}
We evaluate \name in online serving of SWE-Bench using Mooncake arrival timing. Figure~\ref{fig:ablation-online-comparison} shows that \name improves both TTFT and time-to-last-token (TTLT), achieving approximately 1.64$\times$ speedups in both p95 metrics over the fastest baseline. The TTLT timeline further shows that \name maintains lower request latency as queues build, demonstrating that its benefits extend to full-response completion.

\textbf{Impact of batch size.}
Figure~\ref{fig:ablation-batch-size} shows that \name consistently achieves
2.8$\times$--2.9$\times$ speedups in mean TTFT over Cake across batch sizes
(1, 4, and 8 on H100, Llama-3.1-8B). This improvement increases slightly
with larger batch sizes, demonstrating that \name maintains its latency
advantage under batching.

\section{Conclusion}
Efficient restoration of cached model state is becoming essential as LLM
serving shifts toward longer, repeatedly reused contexts.
Modern hybrid architectures further complicate restoration: full-attention
and recurrent layers exhibit different state footprints and recomputation
semantics that are poorly captured by existing strategies.
\name exploits 3D parallelism across tokens, layers, and devices while
coordinating restoration across concurrent requests.
Its staircase abstraction and batch-aware planning enable flexible
compute--I/O partitioning across  diverse models, workloads, and hardware configurations, reducing
TTFT by 2.24$\times$--3.00$\times$ and TTLT by 1.64$\times$.

\subsection*{AI use statement}
We used generative AI tools to assist in implementing parts of our project code and to suggest possible approaches for optimizing the planning procedure.
We manually derived and verified the dynamic programming framework presented in the paper.
We have also reviewed and tested the AI-generated code for correctness.
We reviewed all AI-assisted work and take responsibility for the final content of this work, including all text, claims, and artifacts produced with the aid of generative AI.

\subsection*{Ethics statement}
This work focuses on system-level optimization to accelerate LLM serving.
Our contributions concern computational efficiency rather than changes to model capabilities or behavior.
We do not anticipate specific ethical concerns arising from the proposed optimization methods.

\subsection*{Reproducibility statement}
We provide the dynamic programming formulation, its assumptions, and the planning algorithm in \S\nobreak\ref{sec:planned-partitioning}.
Section~\ref{sec:setup} describes the experimental setup, including hardware and software configurations, evaluated models and workloads, baseline implementations and performance measurement procedures.
We will release the source code upon acceptance.
To facilitate reproducibility, we will include the configurations and scripts used to run our experiments in the code release.

\bibliography{references}
\bibliographystyle{iclr2027_conference}

\appendix
\clearpage

\appendix

\section{Theoretical Analysis}\label{app:theory}
\subsection{Optimality and Approximation Guarantees of DP Planning}\label{app:dp}
 
\paragraph{Overview.}
We bound how far the plan returned by CacheFlow's DP can be from $T^*$, the
minimum restoration time allowed by the model below, including an uncached
suffix (Lemma~1). The DP solves its own problem exactly (Theorems~1--2), but
that problem restricts plans in three ways:
(i)~all layers in a block share one boundary (Theorem~2, at most $R/B$);
(ii)~boundaries lie on the chunk grid (Theorem~3, at most $Ls(C-1)/B$); and
(iii)~transfer is counted in chunk units with rounding error $e$
(Theorem~4, at most $L|e|/B$). Within the model these are the only losses;
Proposition~1 in \S3.2 states them in relative form, and
Appendix~\ref{app:speedup} extends the model to multiple devices. The
analysis does not yet cover sublayer shortcuts, additional snapshots,
heterogeneous blocks, non-additive batched costs, or estimated costs; we
leave these extensions to future work.
 
\paragraph{Model and assumptions.}
We restore one request with $N\geq1$ cached tokens, followed by $N_s\geq0$
uncached suffix tokens, in a serial network of $L\geq1$ identical blocks
$i=1,\ldots,L$. Each block has $q\geq0$ recurrent (linear-attention) layers
$r=1,\ldots,q$ followed by one full-attention layer. The parameters are:
\begin{itemize}
\setlength{\itemsep}{0pt}
\setlength{\parsep}{0pt}
\item $s>0$: KV bytes per token in one block's full-attention layer.
\item $R_r\geq0$: bytes in recurrent layer $r$'s final state after all $N$
tokens, with $R=\sum_rR_r$ per block ($R=0$ when $q=0$).
\item $F(t)$: time to replay the first $t$ tokens through one block, given
its input activations and initial recurrent states; it is identical across
blocks, nondecreasing for integer $0\leq t\leq N+N_s$, and $F(0)=0$.
\item $B>0$: loading bandwidth in bytes per second.
\end{itemize}
Relative to \S3.2, $N=N_c$, $s=S_{\mathrm{attn}}/C$, $R=S_{\mathrm{rec}}$,
and $F,a,d$ are indexed by tokens rather than chunks, e.g.,
$f(j)=F(\min\{jC,N\})$. Network inputs are available and target caches are
loadable from off device, but no internal activation checkpoints or
intermediate recurrent snapshots are available
(Appendix~\ref{app:speedup} adds shard boundary activations). Replay
evaluates complete layers for selected tokens from known initial states,
without sublayer shortcuts. Costs are exact, nonnegative, and additive, and
full-attention token costs are nondecreasing with position; chunking only
restricts boundaries. One compute resource and one independent loading
resource overlap without interference, with sufficient buffers.
 
\paragraph{Uncached suffix.}
Every block recomputes the suffix, whose tokens attend to the whole prefix
however it was restored. With additive costs, one block's suffix prefill thus
takes the plan-independent time $\sigma=F(N+N_s)-F(N)$, the per-block suffix
time of \S3.2. Block~$i$'s suffix prefill starts only after its
full-attention KV for all $N$ prefix tokens and its final recurrent state are
on device, and after block~$i-1$'s suffix prefill. Restoration ends when
block~$L$ finishes its suffix prefill, or, if $N_s=0$, when all target
caches are on device.
 
\paragraph{DP formulation.}
For chunk size $C\geq1$, the allowed boundaries are
$\mathcal G_C=\{0,C,\ldots,\lfloor N/C\rfloor C\}\cup\{N\}$, where $N$
permits a partial final chunk. A plan $\mathbb{J}=(j_1,\ldots,j_L)$ with
$N\geq j_1\geq\cdots\geq j_L\geq0$ recomputes the first $j_i\in\mathcal G_C$
tokens in block~$i$. For boundary $t$, the bytes saved and still loaded are
\[
a(t)=st+\mathbbm{1}_{\{t=N\}}R,\qquad d(t)=s(N-t)+\mathbbm{1}_{\{t<N\}}R,
\]
since only replay to $N$ replaces loading final recurrent states. With
all-load volume $A_0=L(Ns+R)$, the objective is
\[
T(\mathbb{J})=\max\left\{\sum_{i=1}^{L}F(j_i)+L\sigma,\;
                  \frac{1}{B}\sum_{i=1}^{L}d(j_i)+\sigma\right\},
\]
which Lemma~\ref{app:lemma} relates to the makespan. Let $D_i[x]$ be the minimum compute
time of $i$ boundary choices saving exactly $x$ bytes, with $D_0[0]=0$ and
all other initial or unreachable states, including negative indices, set to
$+\infty$. Then
\begin{align*}
D_i[x]&=\min_{t\in\mathcal G_C}\{D_{i-1}[x-a(t)]+F(t)\},\\
T_{\mathrm{DP}}&=\min_{x:D_L[x]<\infty}
                 \max\{D_L[x]+L\sigma,(A_0-x)/B+\sigma\}.
\end{align*}
Backtracking from a minimizing state and sorting the boundaries in
nonincreasing order gives a feasible plan with the same costs, since blocks
are identical.
 
\begin{lemma}[Suffix overlap]
\label{app:lemma}
Every execution with prefix-replay and loading totals $P$ and $Q$ takes at
least $\max\{P+L\sigma,Q+\sigma\}$. For a sorted plan $\mathbb{J}$ whose
loads are issued in block order, with ready suffix work run before remaining
prefix replay, the makespan is at most
$\max\{T(\mathbb{J}),Q(\mathbb{J})/L+L\sigma\}\leq T(\mathbb{J})+Q(\mathbb{J})/L$,
and it equals $T(\mathbb{J})$ when $Q(\mathbb{J})\geq L\sigma$.
\end{lemma}
\begin{proof}
The compute resource performs $P+L\sigma$ work, and the block whose loads
finish last, no earlier than $Q$, still needs its suffix prefill. For the
upper bound, let $Q_k$ be when block~$k$'s loads finish. The compute
resource idles only when all prefix replay is done and the next suffix block
awaits its loads. Without idling, the makespan is $P+L\sigma$. Otherwise,
after its last idle period, ending at some $Q_k$, it prefills only blocks
$k,\ldots,L$ and finishes at $Q_k+(L-k+1)\sigma$. Sorting makes $d(j_i)$
nondecreasing, so $Q_k\leq kQ/L$, and the finish time is at most
$\max\{Q/L+L\sigma,Q+\sigma\}$, its values at $k=1$ and $k=L$. This is at
most $T(\mathbb{J})+Q/L$ because $L\sigma\leq T(\mathbb{J})$, and at most
$Q+\sigma$ when $Q\geq L\sigma$.
\end{proof}
 
\begin{theorem}[Full attention, $C=1$]
For $q=0$, $T_{\mathrm{DP}}$ lower-bounds the makespan of every
complete-layer execution, and the DP plan attains it whenever its loading
time is at least $L\sigma$. In particular, the DP is optimal when $N_s=0$.
\end{theorem}
\begin{proof}
Let $U_i$ be the prefix tokens recomputed in layer~$i$; activation
dependencies give $U_{i+1}\subseteq U_i$. Replacing each $U_i$ by its first
$|U_i|$ positions preserves nesting and transfer volume without increasing
compute, and by Lemma~1 the resulting prefix plan's objective is at most the
original makespan. Conditioning on the last boundary then gives the
recurrence for $D_i[x]$; at fixed $x$, loading is fixed, so minimum compute
suffices, and sorting preserves both costs. This DP argument applies to any
$\mathcal G_C$ and $R\geq0$. Attainment follows from Lemma~1 because the DP
plan is sorted.
\end{proof}
 
\begin{theorem}[Loss from shared block boundaries]
The DP is exact among block-prefix plans. Let $T^*_{\mathrm{block},1}$
and $T^*_{\mathrm{layer},1}$ be the token-level optima with shared block
boundaries and independent layer boundaries, the latter lower-bounding
every complete-layer execution. Then
\[
0\leq T^*_{\mathrm{block},1}-T^*_{\mathrm{layer},1}\leq R/B.
\]
\end{theorem}
\begin{proof}
Block exactness follows from Theorem~1. Without intermediate snapshots,
recurrent replay forms a prefix, so Theorem~1's prefix replacement yields an
optimal layer plan with recurrent boundaries $k_{i,r}$ and full-attention
boundaries $j_i$. For $q>0$,
\[
k_{i,1}\geq\cdots\geq k_{i,q}\geq j_i,
\qquad j_i\geq k_{i+1,1}\quad(i<L).
\]
Replacing each $k_{i,r}$ by $j_i$ keeps all required activations, so the
resulting block plan $\widehat{\mathbb{J}}$ is feasible, uses no more
compute, and leaves full-attention traffic unchanged. It loads an extra
final state only when $k_{i,r}=N>j_i$. Let $b$ be the first block with
$j_b<N$; earlier blocks have $j_i=N$, and later recurrent boundaries are at
most $j_b<N$, so only block~$b$ adds loading, at most $R$ bytes. With the
layer plan's totals $P^*,Q^*$,
\[
T^*_{\mathrm{block},1}\leq T(\widehat{\mathbb{J}})
\leq\max\{P^*+L\sigma,Q^*+\sigma+R/B\}
\leq T^*_{\mathrm{layer},1}+R/B.
\]
The lower bound holds because block plans are also layer plans.
\end{proof}
 
\begin{theorem}[Loss from chunk boundaries]
Let $T^*_{\mathrm{block},C}$ be the exact DP value on $\mathcal G_C$. Then
\[
0\leq T^*_{\mathrm{block},C}-T^*_{\mathrm{block},1}
\leq\frac{Ls(C-1)}B.
\]
\end{theorem}
\begin{proof}
Round a token-optimal block plan $\mathbb{J}^*$ down to the grid,
$\widehat j_i=\max\{g\in\mathcal G_C:g\leq j_i^*\}$, so that
$0\leq j_i^*-\widehat j_i\leq C-1$. Rounding preserves order and cannot
increase compute. Since $N\in\mathcal G_C$, a boundary equals $N$ after
rounding exactly when it did before, so
$d(\widehat j_i)-d(j_i^*)\leq s(C-1)$ and loading grows by at most
$Ls(C-1)/B$. The chunk DP is no worse than $\widehat{\mathbb{J}}$, and the
lower bound follows from $\mathcal G_C\subseteq\mathcal G_1$.
\end{proof}
Strict loss is possible: with $L=1$, $N=C=2$, $N_s=R=0$, $s/B=1$, and
$F(t)=t(t+1)/2$, the token optimum is $1$ at $t=1$, whereas the chunk
optimum is $2$ at $t=0$, attaining the bound.
 
\begin{theorem}[Loss from unit rounding]
Suppose the planner indexes $x$ in units of $S_{\mathrm{attn}}=Cs$ bytes and
rounds $a(N)=sN+R$ to $a(N)+e$, so that $|e|\leq S_{\mathrm{attn}}/2$ under
nearest rounding. Then its plan $\tilde{\mathbb{J}}$ satisfies
$T(\tilde{\mathbb{J}})\leq T^*_{\mathrm{block},C}+L|e|/B$.
\end{theorem}
\begin{proof}
Each $a(t)$ with $t<N$ is already a multiple of $S_{\mathrm{attn}}$, so each
plan's estimated load errs by $e\,\nu(\mathbb{J})/B$ with
$\nu(\mathbb{J})=|\{i:j_i<N\}|\leq L$. The estimated objective $\tilde T$
thus lies in $[T,T+L|e|/B]$ if $e\geq0$ and in $[T-L|e|/B,T]$ otherwise. For
$e\geq0$ and $\mathbb{J}^*_C$ attaining $T^*_{\mathrm{block},C}$,
$T(\tilde{\mathbb{J}})\leq\tilde T(\tilde{\mathbb{J}})\leq\tilde T(\mathbb{J}^*_C)\leq T(\mathbb{J}^*_C)+L|e|/B$;
the case $e<0$ is symmetric.
\end{proof}
 
\paragraph{Magnitude of the approximation terms.}
Write $T^*=T^*_{\mathrm{layer},1}>0$ and let $\kappa>0$ satisfy
$F(t)\geq\kappa t$. A plan recomputing $Z=\sum_ij_i$ full-attention
positions needs at least $\kappa Z$ compute and $s(LN-Z)/B$ loading, and the
suffix only adds to both. With $\gamma=1+s/(B\kappa)$,
\[
T^*\geq\min_{0\leq Z\leq LN}\max\Bigl\{\kappa Z,\frac{s(LN-Z)}B\Bigr\}
=\frac{LNs}{B\gamma},
\]
so dividing each additive term by this bound gives
\[
\varepsilon_R=\frac{R}{BT^*}\leq\frac{\gamma R}{LNs},\qquad
\varepsilon_C=\frac{Ls(C-1)}{BT^*}\leq\frac{\gamma(C-1)}{N},\qquad
\varepsilon_e=\frac{L|e|}{BT^*}\leq\frac{\gamma|e|}{Ns}.
\]
All three are $O(1/N)$ and assume nothing about $F$ beyond
$F(t)\geq\kappa t$.
 
\paragraph{Combined guarantee (Proposition~1).}
Theorems~2--4 give
\[
T^*\leq T(\tilde{\mathbb{J}})\leq T^*+\frac{R+Ls(C-1)+L|e|}B,
\qquad \frac{T(\tilde{\mathbb{J}})}{T^*}\leq1+\varepsilon_R+\varepsilon_C+\varepsilon_e.
\]
Since $T^*$ lower-bounds every complete-layer execution and, by Lemma~1,
$T(\tilde{\mathbb{J}})$ is the DP plan's makespan whenever
$Q(\tilde{\mathbb{J}})\geq L\sigma$, the bounds then compare CacheFlow with
the best execution.
 
\paragraph{Long suffixes.}
If instead the suffix prefill outlasts loading, Lemma~1 allows the makespan
to exceed $T(\tilde{\mathbb{J}})$ by up to $Q(\tilde{\mathbb{J}})/L<\sigma$,
since waiting for early loads delays the suffix chain. The bounds then gain
at most one block's suffix prefill, and the suffix, not restoration,
dominates the time to first token.
 
\paragraph{Qwen3.5-9B at 10\,Gbps.}
Qwen3.5-9B has $L=8$ blocks, each with $q=3$ Gated DeltaNet layers (a
$32\times128\times128$ FP32 state and an $8192\times3$ BF16 convolution
state each) and one full-attention layer (4 BF16 KV heads of dimension 256).
Thus $s=4096$\,B and $R=6{,}438{,}912$\,B; with $B=1.25$\,GB/s and $C=512$,
$R$ rounds to $3S_{\mathrm{attn}}$ and $|e|=147{,}456$\,B. The H100 dense
BF16 peak $\Pi=989$\,TFLOP/s gives the valid bound
$\kappa=2P_{\mathrm{blk}}/\Pi\approx1.75\,\mu$s for
$P_{\mathrm{blk}}\approx8.65\times10^{8}$ parameters per block, so
$\gamma\approx2.87$. Assuming at most 50\% MFU, an empirical assumption
rather than a hardware limit, gives $\kappa\approx3.50\,\mu$s and
$\gamma\approx1.94$. The additive gap is at most $19.5$\,ms either way. The
relative bounds below hold for any $N_s$ and are rounded upward
($1\mathrm{K}=1024$ tokens):
\begin{center}
\begin{tabular}{lr|rrrr|rrrr}
\hline
 & & \multicolumn{4}{c|}{Peak ($\gamma\approx2.87$)}
 & \multicolumn{4}{c}{50\% MFU ($\gamma\approx1.94$)} \\
Context & $N$ & $\varepsilon_R$ & $\varepsilon_C$ & $\varepsilon_e$ & Total
 & $\varepsilon_R$ & $\varepsilon_C$ & $\varepsilon_e$ & Total \\
\hline
$32\mathrm{K}$  & $32{,}768$  & $1.73\%$ & $4.49\%$ & $0.32\%$ & $6.52\%$
 & $1.17\%$ & $3.03\%$ & $0.22\%$ & $4.40\%$ \\
$64\mathrm{K}$  & $65{,}536$  & $0.87\%$ & $2.25\%$ & $0.16\%$ & $3.26\%$
 & $0.59\%$ & $1.52\%$ & $0.11\%$ & $2.20\%$ \\
$128\mathrm{K}$ & $131{,}072$ & $0.44\%$ & $1.13\%$ & $0.08\%$ & $1.63\%$
 & $0.30\%$ & $0.76\%$ & $0.06\%$ & $1.10\%$ \\
\hline
\end{tabular}
\end{center}
 
\subsection{Near-Linear Speedup of Device-wise Parallelism}\label{app:speedup}
\paragraph{Setup.}
We split the $L$ blocks of Appendix~\ref{app:dp} evenly into $S$ contiguous
shards on identical devices, each with loading bandwidth $B$. Every shard
after the first loads a cached boundary activation of $h$ bytes per token
before replay, in place of the network input. The suffix traverses the
shards in order, and we ignore transferring its activations between devices.
Let $T^\star_{\mathrm{single}}$ be the minimum of $T(\mathbb{J})$ over
single-device plans, and $T_{\mathrm{3D}}$ the concurrent restoration time
with per-shard DP plans under the best schedule.
 
\begin{theorem}[Near-linear speedup]
With $\Delta=\max\{F(N),(sN+R)/B\}+hN/B$,
\[
T^\star_{\mathrm{single}}/S\;\leq\;T_{\mathrm{3D}}\;\leq\;
T^\star_{\mathrm{single}}/S+\Delta+(1+1/S)L\sigma.
\]
\end{theorem}
\begin{proof}
Lower bound: let shard~$s$ have prefix-replay and loading totals $P_s,Q_s$,
excluding its boundary activation. It also prefills the suffix of its $L/S$
blocks, and its last load precedes at least one block's suffix prefill, so
$T_{\mathrm{3D}}\geq\max\{P_s+L\sigma/S,\,Q_s+\sigma\}$. Concatenating and
sorting the shard plans gives a single-device plan, so
$T^\star_{\mathrm{single}}\leq\max\{\sum_sP_s+L\sigma,\sum_sQ_s+\sigma\}
\leq S\,T_{\mathrm{3D}}$.
Upper bound: sort an optimal suffix-free single-device plan,
$j_1\geq\cdots\geq j_L$, with totals $P,Q$, so that
$\max\{P,Q\}\leq T^\star_{\mathrm{single}}$, and give shard~$g$ the
boundaries $j_g,j_{g+S},\ldots$. Since $F(j_i)$ is nonincreasing, each term
after the first is at most the mean of the $S$ terms before it, so shard~$g$'s
compute is at most $P/S+F(N)$. Since $d(j_i)$ is nondecreasing, comparing
each term but the last with the $S$ terms after it bounds its loading by
$Q/S+(sN+R)/B$. Shard~$g$'s DP objective, and hence its prefix-restoration
time, is thus at most
$T^\star_{\mathrm{single}}/S+\max\{F(N),(sN+R)/B\}+L\sigma/S$. Loading its
boundary activation first adds at most $hN/B$, and prefilling the suffix
through all $L$ blocks afterwards adds $L\sigma$.
\end{proof}
 
\paragraph{Interpretation.}
$\Delta$ does not grow with $L$, whereas the suffix term is serial because
the suffix traverses every shard. Since
$T^\star_{\mathrm{single}}\geq LNs/(B\gamma)$ (Appendix~\ref{app:dp}),
\[
S\;\geq\;\frac{T^\star_{\mathrm{single}}}{T_{\mathrm{3D}}}
\;\geq\;S\Bigl(1-\frac{S}{L}\cdot\frac{\gamma B\Delta}{Ns}
-(S+1)\frac{\gamma B\sigma}{Ns}\Bigr).
\]
The speedup is near-linear when each shard holds many blocks, boundary
activations are small relative to per-block KV, and a block's suffix prefill
is short compared with loading its prefix KV.

\section{Additional Results}
\label{app:large-moe}

\begin{figure}[htbp]
\captionsetup{hypcap=false}
\centering
\begin{minipage}[t]{0.32\linewidth}
\vspace{0pt}
\centering
\includegraphics[width=\linewidth]{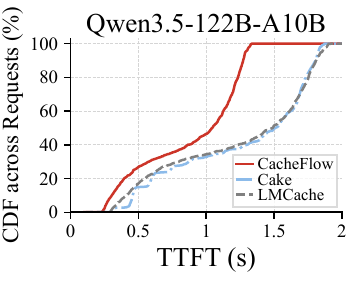}
\caption{Large-scale MoE serving (TP=8, PP=1, EP disabled).}
\label{fig:large-moe-tp8-pp1-epoff}
\end{minipage}
\hfill
\begin{minipage}[t]{0.65\linewidth}
\vspace{0pt}
\centering
\nextfloat
\begin{subfigure}[t]{0.49\linewidth}
    \centering
    \includegraphics[width=\linewidth]{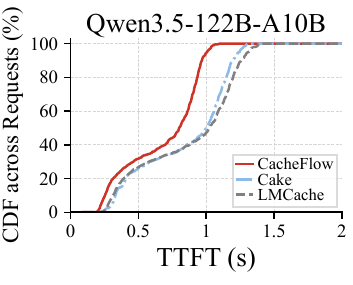}
    \caption{EP disabled.}
\end{subfigure}
\hfill
\begin{subfigure}[t]{0.49\linewidth}
    \centering
    \includegraphics[width=\linewidth]{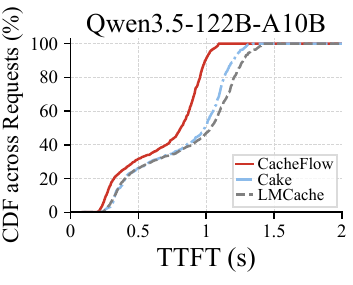}
    \caption{EP enabled.}
\end{subfigure}
\caption{Large-scale MoE serving (TP=4, PP=2).}
\label{fig:large-moe-tp4-pp2}
\end{minipage}
\end{figure}

\end{document}